\documentclass[conference]{IEEEtran}
\IEEEoverridecommandlockouts
\usepackage{cite}
\usepackage{algorithmic}
\usepackage{xcolor}
\usepackage{enumerate}
\usepackage{amsfonts,amsmath,amssymb,amsfonts}
\usepackage{algorithmic}
\usepackage{optidef}
\usepackage{algorithm}
\usepackage{graphicx}
\usepackage{textcomp}
\usepackage{subfigure}
\usepackage[section]{placeins}
\def\BibTeX{{\rm B\kern-.05em{\sc i\kern-.025em b}\kern-.08em
    T\kern-.1667em\lower.7ex\hbox{E}\kern-.125emX}}
\begin{document}
\bibliographystyle{IEEEtran}
\title{Federated Learning for Distributed Energy-Efficient Resource Allocation}
\author{\IEEEauthorblockN{Zelin Ji,~\IEEEmembership{Student Member,~IEEE},
Zhijin Qin,~\IEEEmembership{Senior Member,~IEEE}}
\IEEEauthorblockA{School of Electronic Engineering and Computer Science\\ Queen Mary University of London, London, UK\\
Email: \{z.ji, z.qin\}@qmul.ac.uk}
}
\maketitle

\begin{abstract}
In cellular networks, resource allocation is performed in a centralized way, which brings huge computation complexity to the base station (BS) and high transmission overhead. This paper investigates the distributed resource allocation scheme for cellular networks to maximize the energy efficiency of the system in the uplink transmission, while guaranteeing the quality of service (QoS) for cellular users. Particularly, to cope the fast varying channels in wireless communication environment, we propose a robust federated reinforcement learning (\textit{FRL\_{suc}}) framework to enable local users perform distributed resource allocation in items of transmit power and channel assignment by the guidance of the local neural network trained at each user. Analysis and numerical results show that the proposed \textit{FRL\_{suc}} framework can lower the transmission overhead and offload the computation from the central server to the local users, while outperforming the conventional multi-agent reinforcement learning algorithm in terms of EE, and is more robust to channel variations.

\end{abstract}

\section{Introduction}
The 3rd Generation Partnership Project (3GPP) has designed the access technique standard and physical channel model for fifth generation new radio (5G NR) network \cite{3gpp_38.211, 3gpp_38.901}. Unlike the fixed subcarrier spacing in Long-Term Evolution (LTE), 5G NR supports multiple subcarrier spacing, and the bandwidth part (BWP) technique in 5G NR enables the user equipment (UE) to switch between different resource blocks (RBs) with different bandwidths dynamically. Such new techniques put forward higher requirements on the resource allocation in 5G and beyond network, and how to assign the RBs to improve the overall quality of service (QoS) of systems become an issue that needs to be resolved urgently. Specifically, most cellular UEs are battery-powered, and the rate maximization-oriented algorithms~\cite{1561930} may lead to unnecessary energy consumption, which is adverse to the development of the massive capacity and connectivity trend for 5G and beyond communications.

\subsection{Energy-Efficient Resource Allocation for Cellular Networks}

The literature on energy-efficient resource allocation mainly focuses on transmit power and channel assignment optimization~\cite{7579565, 4205089}. Robat Mili \emph{et al.}~\cite{7579565} maximize energy efficiency (EE) for device-to-device communications. Although there have been a rich body of works investigating the resource allocation in the wireless communication system, most of works are centralized based, which are considered to be complex and not easily scalable~\cite{4205089}. For the scope of resource allocation, the center acquires the global CSI to assign the channel to UEs, leading to huge communication overhead and high communication latency. Therefore, distributed algorithms are preferred than centralized ones.

Game theory has been adopted for the distributed resource allocation~\cite{4205089, 9280358, 8103020}. However, it usually requires several iterations for UEs to converge to the Nash Equilibrium (NE) point, thus requiring the radio environment to be static. In practical environment, the fast-varying channel will affect the performance of game theory based algorithms. Yang \emph{et al.} \cite{9280358} and Dominic \emph{et al.} \cite{8103020} combine the game theory and stochastic learning algorithm (SLA) to enable local users learn from the previous experience and cope the channel variations. However, the game theory based algorithms fail to explore the benefits of collaboration and communication between users, which may affect the system level performance.

\subsection{Machine Learning Algorithms for Resource Allocation}
Based on the discussion above, a promising solution is to establish a decentralized resource allocation framework and extend the intelligent algorithm to a cooperative large-scale network. The multi-agent reinforcement learning (MARL) algorithm shows the opportunity to cope the complexity challenge and enhance the intelligence of the local nodes. MARL algorithms only relay on the real-time local UEs information and observations, thereby reducing the communication overhead and the latency significantly. MARL approaches have been widely applied in wireless communications~\cite{liang2019spectrum,9026965}. Wang \emph{et al.} \cite{9026965} have verified that such decentralized optimization approach can achieve near-optimal performance. The main challenge of MARL algorithms is the unstable and unpredictable actions of other UEs lead to an unstable environment, which makes MARL hard to convergence~\cite{foerster2017stabilising}. To overcome the non-stationary challenge, some essential information sharing between UEs needs to be guaranteed, which could be supported by federated learning (FL)~\cite{9530714}. FL enables the local users, to collectively train a global model using their raw data while keeping these data locally stored on the mobile devices~\cite{8994206}, and has been successful applied in the next-word prediction~\cite{hard2019federated} and the system level design~\cite{bonawitz2019federated}. Particularly, federated reinforcement learning (FRL) enables UEs explore the environment individually, and benefit from others' experience by training a global model collaboratively. Compared with MARL approaches, the FRL method enables the UEs to transfer the experience with each other thus improves the convergence performance~\cite{45895}. Inspired by this, Zhang \emph{et al.}~\cite{9348485} and Zhong \emph{et al.}~\cite{zhong2021mobile} have applied FRL for enhancing WiFi multiple access performance and optimizing the location of reconfigurable intelligent surface. 


All of such valuable works encourage us to apply the FRL to the channel assignment and power optimization  problems to offload the computation to local UEs, to reduce the transmission overhead, and to enable collaboration among UEs. In this paper, we apply a FRL based framework to solve the channel assignment and power optimization problem distributedly. To the best of our knowledge, this is the first work to apply the FRL framework for the resource allocation problem. The contribution of this paper is concluded as follows:
\begin{enumerate}[1)]
\item {A FRL framework, named \textit{FRL\_{suc}}, is proposed to jointly optimize the channel assignment and transmit power. The optimization is performed distributed at local UEs to lower the computational cost at the BS and the transmission overhead. }

\item {To explore the collaboration among cellular users, we propose to adopt a global reward for all UEs and apply the \textit{FRL\_{suc}} framework for experience sharing among UEs.}

\item {To improve the success rate of the resource allocation, a novel averaging algorithm is proposed for the local model averaging process. The models are averaged according to the success rate of the channel allocation to improve the performance of the proposed \textit{FRL\_{suc}} framework.}

\end{enumerate}
The remainder of the paper is organized as follows. In Section~\uppercase\expandafter{\romannumeral2}, the system model is presented and an EE maximization problem is formulate. {The proposed FRL framework is presented and the communication cost of it is analyzed in Section~\uppercase\expandafter{\romannumeral3}.} The numericafl results are illustrated in Section~\uppercase\expandafter{\romannumeral4}. The conclusion is drawn in Section~\uppercase\expandafter{\romannumeral5}.

\section{System Model}

In this paper, we consider a system model with orthogonal frequency-division multiple access (OFDMA). It is assumed that the set of UEs is denoted as ${\cal UE} = \left \{UE_1, \dots, UE_I \right\}$, where $I$ is the total number of UEs. For $UE_i$, the binary channel assignment vector is given by $\boldsymbol {\rho}_i = \left [\rho_{i,1}, \dots,\rho_{i,n}, \dots, \rho_{i,N} \right], i \in I, n\in N$, where $N$ is the number of sub-channels, $\rho_{i,n}=1$ indicates that the $n$-th sub-channel is allocated to $UE_i$, otherwise $\rho_{i,n} = 0$. Each UE is limited to access only one channel, i.e., $\sum \nolimits ^N_{n=1} \rho_{i,n} = 1, \forall i \in I$. Meanwhile, a channel can be accessed by at most one UE with in a cluster, i.e., $\sum \nolimits^I_{i=1} \rho_{i,n} \in \{0,1\} , \forall n \in N$. For each UE, if it access one sub-channel and there are no other UEs access this sub-channel within a cluster, then it can transmit data with the BS successfully. If each UE is assigned with a channel without conflict with any other UE within the same cluster, then the assignment is defined as a successful channel assignment. The success rate for the channel assignment can be defined as $\eta_i = {\xi}_i/{T}$, where ${\xi}_i$ represents the successful assignment counts for $UE_i$ in the past, and $T$ represents the number of resource allocation counts since the initialization the system.

The pathloss model between $UE_i$ and the BS can be denoted by~\cite{3gpp_38.901}
\begin{equation}
PL_{i,n} = 32.4 + 20 \log_{10}\left (f_n \right) + 30 \log_{10} \left (d_{i,n}\right) (\text{dB}),
\label{eq1}
\end{equation}
where $f_n$ represents the carrier frequency for $n$-th sub-channel, $d_{i,n}$ represents the 3D distance between $UE_i$ and the BS. The overall channel gain can be thereby denoted by
\begin{equation}
    h_{i,n} = \underbrace{\frac{1}{10^{(PL_{i,n}/10)}} \psi}_{\text{Large scale fading}} m_n,
\label{eq2}    
\end{equation}
where $\psi$ is the shadowing parameter, which is log-normally distributed. We assume that there is no line of sight link between UEs and the BS, and $m_n$ represents the Rayleigh fading power component of the $n$-th sub-channel.

The signal-to-noise ratio (SNR) between the BS and $UE_i$ transmitting over the $n$-th sub-channel is represented as 
\begin{equation}
    \gamma_{i,n} = \frac{\rho_{i,n}h_{i,n}p_{i}}{N_n},
\label{eq3}
\end{equation}
where $N_n = \sigma_n^2$ represents the Gaussian noise power on the $n$-th sub-channel. The uplink EE for a successful channel assignment of $UE_i$ on the $n$-th sub-channel is given by 
\begin{equation}
u_{i,n} =
\begin{cases}
\frac{BW_n}{p_i} \log_2\left (1+\gamma_{i,n}\right), &\text{if  } \sum \nolimits ^I_{i=1} \rho_{i,n} = 1;\\
0,&\text{else}.
\end{cases}
\label{eq4}
\end{equation}
where $BW_n = k\times b_n$ is the bandwidth of the $n$-th sub-channel, $k$ represents the number of subcarriers in each sub-channel, and $b_n$ denotes the subcarriers spacing for $n$-th sub-channel. Meanwhile, for the unsuccessful assignment, i.e., there are more than one UE access one sub-channel at the same time, the uplink rate is set to $0$ since it is unacceptable for the OFDMA system.



The problem is formulated as
\begin{maxi!}|l|
{\{\boldsymbol \rho, \boldsymbol p\}}{\sum^I_{i=0} \sum^N_{n=0} u_{i,n}}
{\label{eq5}}{\text{P0:}}
\addConstraint{ p_i \leq p_{max}, \forall i \in I \label{objective:c1} }
\addConstraint{\gamma_{i,n} > \gamma_{min}, \forall i \in I \label{objective:c2} }
\addConstraint{\sum ^N_{n=1} \rho_{i,n} = 1
, \forall i \in I \label{objective:c3} }
\addConstraint{\sum \nolimits ^I_{i=1} \rho_{i,n} \in \{0, 1\}, \forall n \in N. \label{objective:c4} }
\end{maxi!}
where $\boldsymbol p = \left \{p_1, \dots, p_I \right\}$ denotes the transmit power vector of UEs, $\gamma_{min}$ represents the minimum SNR requirement to guarantee the QoS for UEs. Constraint (\ref{objective:c3}) and (\ref{objective:c3}) makes the EE maximization problem a non-convex optimization problem and cannot be solved by mathematical convex optimization tools. In the previous work, the channel allocation problems are usually formed as a linear sum assignment programming (LSAP) problem. To solve this problem, local CSI or the UE related information, e.g., location and velocity need to be uploaded to the BS, then the centralized Hungarian algorithm~\cite{3800020109} can be invoked to solve the problem with computational complexity $O\left (I^3\right)$, which means high transmission overhead and lead to high computational pressure to the BS. Furthermore, the mobility of UEs causes the unstable CSI, which means the optimized resource allocation scheme by the BS may not be optimal for current UEs any more. To overcome these challenges, a distributed resource allocation approach is more than desired.


\section{Proposed Federated Reinforcement Learning for Resource Allocation}
In this section, we will first introduce the basis of RL and then propose a FRL framework to enable the distributed resource allocation for solving (P0).

\subsection{Preliminaries of Multi-Agent Reinforcement Learning and Deep Q-network}
{
MARL enables local UEs to interact with the environment individually and learn the optimal strategy by trial-and-error interaction with the fast-varying environment. Mathematically, the MARL can be modeled as an markov decision process (MDP), and at each training step, the training agents observe the environment state and determine an action based on the current policy. The corresponding reward is received by agents which evaluates the immediate effect of the given state and action pair. Then policy updates based on the received reward and the specific state and action pair. Finally, the environment turns to a new state.}

However, the local UEs cannot directly acquire the global environment state, and if the UEs are not aware of the policy of other UEs, they may choose a channel that has been occupied by another UE, which leads to transmission failure in the OFDMA scheme. As a consequence, a partly collaborative MARL structure with communication among UEs is required. Specifically, each agent can share its reward, RL model parameters, action, and state with other agents. For different collaborative RL algorithms, they may share different RL information. For example, some collaborative MARL algorithms require the agents to share their state and action information, and some require the agents to share their rewards. The training complexity and performance of a collaborative MARL algorithm depend on the data size that each agent needs to share. 

The issue becomes severer when we combine the neural network (NN) with reinforcement learning. Traditional centralized deep Q-network (DQN) stores the environment interactive experience, i.e., transition, to the replay memory and uses these data to train the DQN model. However, in the multi-agent DQN, the local observation cannot represent the global environment state, thus reducing the effectiveness of the replay memory significantly. Although some works have been proposed to enable replay memory for MARL, the solution lacks scalability and cannot make the good tradeoff between the communication cost and performance.

\subsection{Proposed Federated Reinforcement Learning Framework}
\begin{figure}[t]
\centering
\includegraphics[width=\columnwidth, height = 0.8\columnwidth]{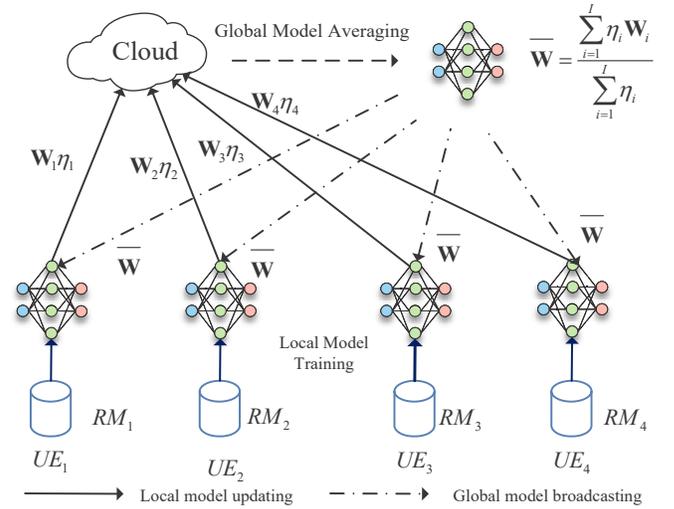}
\centering
\centering
\caption{The proposed federated learning model, where $RM_i$ represents the local replay memory of $UE_i$. The local models are trained using the local replay memory individually, and the local models are uploaded and averaged periodically.}
\label{FL_framework}
\end{figure}
All of these challenges motivate us to design an architecture, where local models can be trained using a local database, i.e., local CSI, and the local models can be united as a global model so that the UEs could learn the knowledge from other UEs experiences and improve the success rate. One popular way is to average the distributed models and form a global model, which is called federated learning (FL)~\cite{9530714}. {As illustrated in Fig. \ref{FL_framework}, a success rate based federated reinforcement learning (\textit{FRL\_{suc}}) framework is designed to optimize the channel assignment and transmit power of UEs, and to improve the success rate and the EE of the system. Compared with supervised learning which requires sufficient data set and pre-knowledge of the system, the (\textit{FRL\_{suc}}) framework can train the local model with the local CSI data which is required by interacting with the environment, thus not only offloading the computational pressure to the UEs, but also lower the transmission overhead significantly.}

\subsubsection{MARL components}
{In this paper, we define the observation state at training step $t$ for the UEs, which are considered as the agents in the \textit{FRL\_{suc}} framework,  as $o_{t, i} = \{\{h_{i,n}\}_{\forall n \in N}, ep, \epsilon\}$ with dimension~$|o_i|$, where $ep$ represents the number of the training epochs and the parameter $\epsilon \in [0,1]$ denotes the probability to explore the new action. The variables $t$ and $\epsilon$ can be treat as a low-dimensional \textit{fingerprint} information to contain the policy of other agents~\cite{foerster2017stabilising}, thus enhance the stationary and the convergence performance of the MARL algorithm. }

The action $a_{t, i}$ for the $UE_i$ including the sub-channel and the transmit power choice with dimension $|a| = 2$. Assuming that the UE can choose the transmit power from a set consists of $N_p$ multiple discrete levels, and the dimension of the action space for each UE can be there by expressed by $|A| = N_p \times N$. The joint action $\boldsymbol{a}_t = \{a_{t, 1}, \dots, a_{t, I}\}$ includes the action choice of all UEs. Note that for the centralized RL algorithm, the dimension of the action space is enlarged to $|A^{\text{cent}}| = (N_p \times N)^I$.

Since we aim to maximize the sum EE of the cellular network, here we design a global reward $r_t$, according to the joint action $\boldsymbol{a}_t$ such that encouraging collaboration of UEs. The global reward at training step $t$ can be defined as
\begin{equation}
r_t=
\begin{cases}
\rho \sum^I_{i=0} \sum^N_{n=0} u_{i,n}(t), &\text{if (\ref{objective:c2}) and (\ref{objective:c4}) are satisfied};\\
0,&\text{Otherwise},
\end{cases}
\label{eq6}
\end{equation}
where $\rho$ is a constant coefficient.
{The objective of the proposed \textit{FRL\_{suc}} framework is to enable UEs to learn a strategy that maximize the expected discount reward, which can be expressed by 
\begin{equation}
R_t = \sum^{\infty}_{\tau=0}\gamma^\tau r_{t+\tau},
\label{eq7}
\end{equation}
where $r_{t+\tau}$ is the immediate reward at $(t+\tau)$-th training step, $\gamma \in (0, 1)$ represents the discount rate, which denotes the impact of the future reward to the current action.}

\subsubsection{Training process for the local model}
The local models are trained using the data in local data set, which is called the replay memory in FRL. At each training step $t$, the experience $e_{t, i} = (o_{t, i}, a_{t, i}, r_t, o_{t+1, i})$ acquired by $UE_i$ is stored in $i$-th local replay memory ${\cal RM}_i$. The optimal policy can be expressed by $\pi_i^*=\arg\max_{{\pi_i}}q^{\pi_i} (o_i,a_i)$, where the policy $\pi$ is defined as a mapping from the observation $o_i$ to the probability of choosing each action, and $q^{\pi_i}(o_i,a_i)= \mathbb{E}_{\pi_i}[R_t|o_{t, i}=o_i, a_{t, i}=a_i]$ represents the expectation discount reward for a state-action pair $(o_i,a_i)$. By learning from the optimal policy $\pi_i^*$, we can find the maximized Q-value~\cite{mnih2013playing}
\begin{equation}
q^{\pi_i^*}\left (o_{t, i}, a_{t, i}\right)= \\
\mathbb{E}\left [r_t+\gamma\max\limits_{a'_{t, i} \in |A|} q^{\pi_i^*}\left (o_{t+1, i},a'_{t, i}\right )|o_{t, i}, a_{t, i}\right],
\label{eq9}
\end{equation}
where $a'_{t, i}$ represents the action that can achieve maximized expectation discount reward for the observation $o_{t+1, i}$. In fact, the UEs can approach to the optimal policy, i.e., $q(o_{t, i}, a_{t, i}) \rightarrow q^{\pi_i^*}(o_{t, i}, a_{t, i})$ as $t \rightarrow \infty$~\cite{sutton2018reinforcement}, but it is impractical since learning steps are discrete. The idea of DQN is to estimate the action-value function by applying the NNs, i.e., $q(o_{t, i}, a_{t, i}; \boldsymbol W) \approx q^{\pi_i^*}(o_{t, i}, a_{t, i})$, where $\boldsymbol W$ represents the weights of the NN in DQN~\cite{mnih2013playing}. Particularly, DQN can deal with large state and action space, which is the main challenge for conventional RL as it need to store and maintain each state-action pair in the large Q-table. At each training step, a minibatch $e_{s,i} = (o_{s, i}, a_{s, i}, r_s, o_{s+1, i})$ is sampled from the replay memory, for which the target Q-value is defined as
\begin{equation}
q_{target}=r_s+\gamma q\left (o_{s, i},\mathop{\arg\max}\limits_{a'_i\in |A|} q\left (o_{s+1, i},a';\boldsymbol W\right);\boldsymbol W^-\right),
\label{eq10}
\end{equation} 
where $\boldsymbol W$ and $\boldsymbol W^-$ denote the parameters of the evaluation network and the target network, respectively. The weights in the evaluation network are updated to the target network periodically. The training loss for the $i$-th local model is denoted by the mean-square error of the target Q-value and the evaluate Q-value, which is expressed by 
\begin{equation}
L(\boldsymbol W_i)=\mathbb E\left [(q_{target}-q(o_{s, i},a_{s, i};\boldsymbol{W_i}))^2\right].
\label{eq17}
\end{equation}
Then weights of the local model are optimized by the gradient descent method~\cite{mnih2013playing}. 

\subsubsection{Averaging of local models}
The individual models at each UE can be united by federated learning. The local model is averaged to a global model, then the global model is broadcast to UEs and the UEs will continue to train the new global model locally. By averaging the models, each UE is able to benefit from the experience of other UEs, since the weights is the direct corresponding to the experience and memory. Mathematically, the model averaging process at the central BS can be denoted as

\begin{equation}
\overline{\boldsymbol W} = \frac{\sum^I_{i=1} |{\cal RM}_i| \boldsymbol W_i}{\sum^I_{i=1} |{\cal RM}_i| },
\label{eq11}
\end{equation}
where $|{\cal RM}_i|$ represents the number of number of elements in ${\cal RM}_i$. The average algorithm shows that the averaged model will learn more from the model with more training cases. However, in the proposed \textit{FRL\_{suc}} framework, we assume that UEs share the a team stage reward, which means the replay memory of each UE has equivalent size. To ensure that the averaged model can benefit from the model that satisfy the QoS requirement, we further revised the averaging algorithm that consider the success rate, which is denoted by
\begin{equation}
\overline{\hat{\boldsymbol W}} = \frac{\sum^I_{i=1}\eta_i \boldsymbol W_i}{\sum^I_{i=1}\eta_i}.
\label{eq12}
\end{equation}

\subsection{Communication Cost Analysis}
In the centralized RL algorithm, UEs upload their local observations information to the BS, and execute the action obtained from the BS. The uplink communication cost is given by
\begin{equation}
C_{\text{CRL}} = K \times T \times \sum^I_{i = 0}|o_i|,
\label{eq13}
\end{equation}
where $K$ and $T$ represents the number of maximum training epochs and the number of training steps in each training epoch. The fast fading channel and large scale fading channel are renewed in each training step and epoch, respectively. Assuming that the fast fading is updated every 1ms, while the pathloss and shadowing are updated every 100ms, which means we also set $T = 100$ training steps in each training epoch. While in the proposed \textit{FRL\_{suc}} framework, UEs only receive the global reward at each training step. The main communication cost comes from the model averaging stage.\footnote{In this paper, we assume that the model can be perfectly uploaded and broadcast, and the failure of transmission of NN models is beyond the scope of this article. } Corresponding uplink communication cost can be given by
\begin{equation}
C_{\text{FRL}} = M \times \sum^I_{i = 0}(|\boldsymbol W_i|+1),
\label{eq14}
\end{equation}
Where $M$ represents the model averaging times, $|\boldsymbol W_i|$ represents the size of the $i$-th local model. Note that the model averaging is less frequent for the \textit{FRL\_{suc}} algorithm than the traditional FL algorithm since the target Q network also needs to be updated periodically. Moreover, the model average period should be an integer multiple of the model update period. We can see that the communication cost for \textit{FRL\_{suc}} is much lower than a centralized algorithm when the model averaging frequency is low.

\section{Numerical Results}

{We consider a communication scenario underlying a single cellular network. The UEs are distributed in a 100m×100m square with a random starting location, and the BS is fixed at the center of the square. We adopt the simulation assumptions in~\cite{3gpp_38.901} to model the channels. The number of UEs and sub-channels are set to $I = 4$ and $N = 4$, respectively. The subcarrier spacing of the four sub-channels are set to 15kHz, 30kHz, 60kHz, and 120kHz, respectively. Each sub-channel contains $k = 12$ subcarriers. To enable the mobility of UEs, we assume that the UEs can move with the speed from 0 meters per second (m/s) to 5 m/s within the square.} We suppose that UEs can choose the discrete transmit power from 0 dBm to 24 dBm with the interval of 3dBm. The noise power is set to $\sigma^2_n = -100$dBm, $\forall n \in N$.



The local model for each UE is made up of an input layer, 3 fully connected hidden layers, and an output layer, containing $|o_k|$, 512, 256, 128, and $|A|$ neurons, respectively. The learning rate for the multi-agent RL models are set to 0.0001, while the $\epsilon$-greedy algorithm is adopted to strike a tradeoff between the exploit and explore. The explore rate $\epsilon$ is set to linearly annealed from 1 to 0.02 over the beginning 4000 epoch and reminds constant afterwards to guarantee the convergence of models.

\begin{figure}[t]
\centering
\includegraphics[width=0.9\columnwidth]{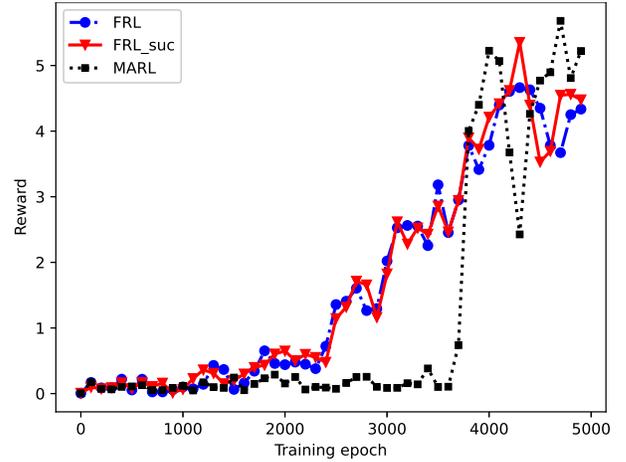}
\caption{{Training performance comparison of the proposed \textit{FRL\_{suc}} algorithm and benchmarks. The model averaging times is set to $n_a=8$. The reward curve shows the averaged reward per 100 steps to enhance the readability.}}
\label{Training_performance}
\end{figure}

\begin{figure}[t]
\centering
\includegraphics[width=0.9\columnwidth]{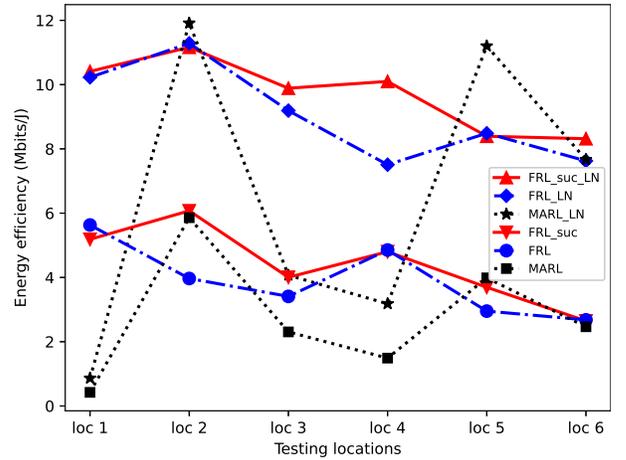}
\caption{{Testing results for 5 random user distributions.  \textcolor[rgb]{0.00,0.00,0.00}{Curves with LN represents the performance in low noise power scenario, i.e., $\sigma_{LN}^2 = -110$dBm.}}}
\label{Test_rate}
\end{figure}

Fig. \ref{Training_performance} illustrates the reward comparison of the proposed \textit{FRL\_{suc}} algorithm, \textcolor[rgb]{0.00,0.00,0.00}{the FRL algorithm, and the traditional MARL algorithm~\cite{liang2019spectrum}}. Due to the punishment, the reward for all schemes is low at the beginning of the training period. With the decrements of the exploration rate $\epsilon$, the proposed \textit{FRL\_{suc}} and the conventional FRL algorithm can achieve faster convergence and stable training reward, while the conventional multi-agent RL algorithm needs more iterations to find the actions with higher reward and converge. \textcolor[rgb]{0.00,0.00,0.00}{Fig. \ref{Test_rate} shows the testing performance for 6 random user distributions. To verify the robustness of the proposed schemes, we test the models under different noise power. We adopt the same scenario with a lower noise power scenario (LN) where the noise power is set to $-110$dBm.  The proposed \textit{FRL\_{suc}} algorithm outperforms other reinforcement learning benchmarks in terms of averaged system EE.} On the other hand, the MARL algorithm behaves worst in majority of situations, while can achieve near-optimal performance in some particular testing epochs. The reason for the poor performance of the MARL lies in the isolation of the local models. Although MARL can achieve almost the same reward with other benchmarks, the local models keep the personality and cannot be robust to the different testing environments. 





\begin{figure}[h]
\centering
\includegraphics[width=0.9\columnwidth]{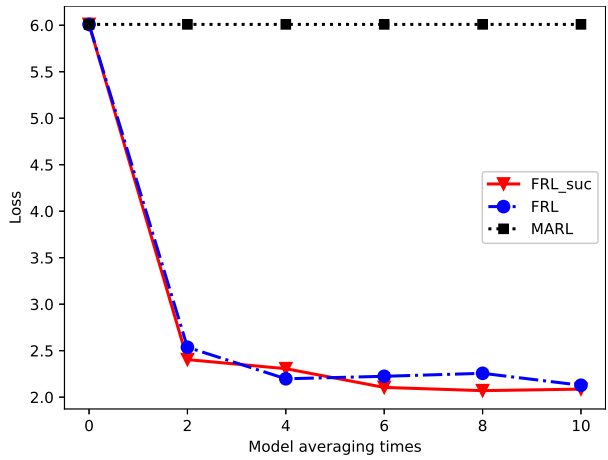}
\centering
\caption{{Training loss over the number of model averaging times.}}
\label{Loss_times}
\end{figure}

\begin{figure}[h]
\centering
\includegraphics[width=0.9\columnwidth]{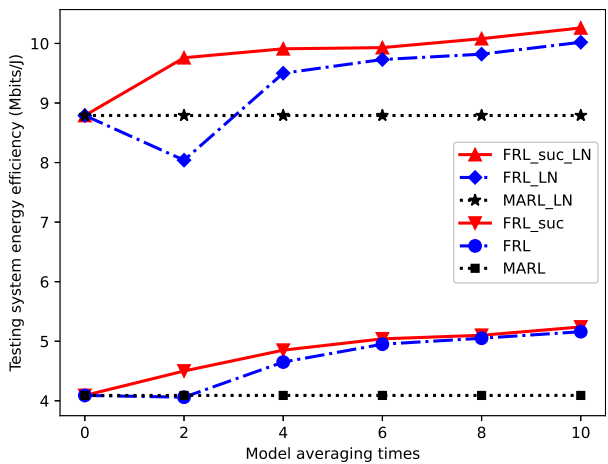}
\centering
\caption{Testing averaged EE performance of 100 random user distributions over the number of model averaging times.}
\label{EE_times}
\end{figure}

Fig. \ref{Loss_times} shows the training loss over the model averaging times. We can see that the MARL algorithm has high training loss even after enough training epochs. As discussed above, the MARL suffers from non-stationary and lacks communication between local models. It is clear that model averaging can decrease the training loss of the local models, representing that local models can benefit from the communication with each other. 


Fig. \ref{EE_times} shows the EE testing performance over the model averaging times. The testing system EE for the proposed \textit{FRL\_suc} algorithm increases with the number of model averaging times grow, and outperforms the MARL scheme when the number of model averaging times is greater than 2. Note that the model averaging at the first several periods may leads the the decrease of the system EE. This is because the averaged model not only loses the personality of the local model, but also neglects the paradigm of the high-performance models. Hence, a well-performed, effective and robust global model requires a few model averaging times. On the other hand, the communication cost also increases linearly with the model averaging times. In practice, it is worth determining the model averaging times according to the requirement in a specific scenario, to strike a tradeoff between the communication cost and the system EE performance.


\section{Conclusion}
In this paper, a distributed energy-efficient resource allocation scheme has been investigated. The system energy efficiency has been maximized by jointly optimize the channel assignment and the transmit power of the user equipments. The formed non-convex problem has been solved by the proposed success rate based federated reinforcement learning framework to overcome the challenge of the computational complexity at the base station and the transmission cost by the local data. Quantity analysis and numerical results have shown that the proposed outperforms other decentralized benchmarks, while lowering the communication cost and offloading the computational complexity from the BS compared to the centralized algorithm. Additionally, the effectiveness of the proposed framework has been verified by illustrating the loss and performance over model averaging times. 


\bibliography{Reference}

\begin{thebibliography}{10}
\providecommand{\url}[1]{#1}
\csname url@samestyle\endcsname
\providecommand{\newblock}{\relax}
\providecommand{\bibinfo}[2]{#2}
\providecommand{\BIBentrySTDinterwordspacing}{\spaceskip=0pt\relax}
\providecommand{\BIBentryALTinterwordstretchfactor}{4}
\providecommand{\BIBentryALTinterwordspacing}{\spaceskip=\fontdimen2\font plus
\BIBentryALTinterwordstretchfactor\fontdimen3\font minus
  \fontdimen4\font\relax}
\providecommand{\BIBforeignlanguage}[2]{{%
\expandafter\ifx\csname l@#1\endcsname\relax
\typeout{** WARNING: IEEEtran.bst: No hyphenation pattern has been}%
\typeout{** loaded for the language `#1'. Using the pattern for}%
\typeout{** the default language instead.}%
\else
\language=\csname l@#1\endcsname
\fi
#2}}
\providecommand{\BIBdecl}{\relax}
\BIBdecl

\bibitem{3gpp_38.211}
\emph{Technical Specification Group Radio Access Network; NR; Physical channels
  and modulation; (Release 16)}, document 3GPP TS 38.211 V16.6.0, 3rd
  Generation Partnership Project, Jun. 2021.

\bibitem{3gpp_38.901}
\emph{Technical Specification Group Radio Access Network; Study on channel
  model for frequencies from 0.5 to 100 GHz (Release 16)}, 3GPP TR 38.901
  V16.1.0, 3rd Generation Partnership Project, Dec. 2019.

\bibitem{1561930}
G.~Song and Y.~Li, ``Utility-based resource allocation and scheduling in
  {OFDM}-based wireless broadband networks,'' \emph{{IEEE} Commun. Mag.},
  vol.~43, no.~12, pp. 127--134, Dec. 2005.

\bibitem{7579565}
M.~{Robat Mili}, P.~{Tehrani}, and M.~{Bennis}, ``Energy-efficient power
  allocation in {OFDMA D2D} communication by multiobjective optimization,''
  \emph{{IEEE} Wireless. Commun. Lett.}, vol.~5, no.~6, pp. 668--671, Dec.
  2016.

\bibitem{4205089}
F.~Meshkati, H.~V. Poor, and S.~C. Schwartz, ``Energy-efficient resource
  allocation in wireless networks,'' \emph{{IEEE} Signal Processing Mag.},
  vol.~24, no.~3, pp. 58--68, May 2007.

\bibitem{9280358}
L.~Yang, D.~Wu, C.~Yue, Y.~Zhang, and Y.~Wu, ``Pricing-based channel selection
  for {D2D} content sharing in dynamic environments,'' \emph{IEEE Trans.
  Wireless Commun.}, vol.~20, no.~4, pp. 2175--2189, Dec. 2021.

\bibitem{8103020}
S.~Dominic and L.~Jacob, ``Distributed resource allocation for {D2D}
  communications underlaying cellular networks in time-varying environment,''
  \emph{IEEE Commun. Lett.}, vol.~22, no.~2, pp. 388--391, Nov. 2018.

\bibitem{liang2019spectrum}
L.~Liang, H.~Ye, and G.~Y. Li, ``Spectrum sharing in vehicular networks based
  on multi-agent reinforcement learning,'' \emph{{IEEE} J. Sel. Areas Commun.},
  vol.~37, no.~10, pp. 2282--2292, Aug. 2019.

\bibitem{9026965}
L.~{Wang}, H.~{Ye}, L.~{Liang}, and G.~Y. {Li}, ``Learn to compress {CSI} and
  allocate resources in vehicular networks,'' \emph{{IEEE} Trans. Commun.},
  vol.~68, no.~6, pp. 3640--3653, Mar. 2020.

\bibitem{foerster2017stabilising}
J.~Foerster \emph{et~al.}, ``Stabilising experience replay for deep multi-agent
  reinforcement learning,'' in \emph{Proc. Int. Conf. Mach. Learn.}, Jul. 2017,
  pp. 1146--1155.

\bibitem{9530714}
Z.~Qin, G.~Ye~Li, and H.~Ye, ``Federated learning and wireless
  communications,'' \emph{IEEE Wireless Commun.}, pp. 1--7, Sep. 2021.

\bibitem{8994206}
J.~Kang, Z.~Xiong, D.~Niyato, Y.~Zou, Y.~Zhang, and M.~Guizani, ``Reliable
  federated learning for mobile networks,'' \emph{{IEEE} Wireless Commun.},
  vol.~27, no.~2, pp. 72--80, Apr. 2020.

\bibitem{hard2019federated}
A.~Hard, K.~Rao, R.~Mathews, S.~Ramaswamy, F.~Beaufays, S.~Augenstein,
  H.~Eichner, C.~Kiddon, and D.~Ramage, ``Federated learning for mobile
  keyboard prediction,'' \emph{arXiv preprint arXiv: 1811.03604}, Feb. 2019.

\bibitem{bonawitz2019federated}
K.~Bonawitz, H.~Eichner, W.~Grieskamp, D.~Huba, A.~Ingerman, V.~Ivanov,
  C.~Kiddon, J.~Konečný, S.~Mazzocchi, H.~B. McMahan, T.~V. Overveldt,
  D.~Petrou, D.~Ramage, and J.~Roselander, ``Towards federated learning at
  scale: System design,'' \emph{arXiv preprint arXiv: 1902.01046}, Mar. 2019.

\bibitem{45895}
H.~H. Zhuo, W.~Feng, Y.~Lin, Q.~Xu, and Q.~Yang, ``Federated deep reinforcement
  learning,'' \emph{arXiv preprint arXiv: 1901.08277}, Feb 2020.

\bibitem{9348485}
L.~Zhang, H.~Yin, Z.~Zhou, S.~Roy, and Y.~Sun, ``Enhancing {WiFi} multiple
  access performance with federated deep reinforcement learning,'' in
  \emph{\textit{Proc.} IEEE Veh. Technol. Conf.}, Nov. 2020, pp. 1--6.

\bibitem{zhong2021mobile}
R.~Zhong, X.~Liu, Y.~Liu, Y.~Chen, and Z.~Han, ``Mobile reconfigurable
  intelligent surfaces for {NOMA} networks: Federated learning approaches,''
  \emph{arXiv preprint arXiv:2105.09462}, Mar. 2021.

\bibitem{3800020109}
\BIBentryALTinterwordspacing
H.~W. Kuhn, ``The hungarian method for the assignment problem,'' \emph{Naval
  Research Logistics Quarterly}, vol.~2, no. 1‐2, pp. 83--97, 1955. [Online].
  Available:
  \url{https://onlinelibrary.wiley.com/doi/abs/10.1002/nav.3800020109}
\BIBentrySTDinterwordspacing

\bibitem{mnih2013playing}
V.~Mnih, K.~Kavukcuoglu, D.~Silver, A.~Graves, I.~Antonoglou, D.~Wierstra, and
  M.~Riedmiller, ``Playing atari with deep reinforcement learning,'' in
  \emph{Proc. NIPS Deep Learn.Workshop}, Dec. 2013.

\bibitem{sutton2018reinforcement}
R.~S. Sutton and A.~G. Barto, \emph{Reinforcement learning: An
  introduction}.\hskip 1em plus 0.5em minus 0.4em\relax MIT Press, 2018.

\end{thebibliography}

\end{document}